\documentclass[aps,showpacs,amsmath,amssymb,prl,superscriptaddress,twocolumn,floatfix]{revtex4-1}
\usepackage{graphicx}
\usepackage{amsmath}
\usepackage{hyperref}
\usepackage{dcolumn}
\usepackage{bm}
\usepackage{color}

\begin{document}

\title{Invariants of broken discrete symmetries}

\author{P.~A.~Kalozoumis}
% \email[]{pkalozoum@phys.uoa.gr}
\affiliation{Department of Physics, University of Athens, GR-15771 Athens, Greece}

\author{C.~Morfonios}
% \email[]{christian.morfonios@physnet.uni-hamburg.de}
\affiliation{Zentrum f\"ur Optische Quantentechnologien, Universit\"{a}t Hamburg, Luruper Chaussee 149, 22761 Hamburg, Germany}

\author{F.~K.~Diakonos}
% \email[]{fdiakono@phys.uoa.gr}
\affiliation{Department of Physics, University of Athens, GR-15771 Athens, Greece}

\author{P.~Schmelcher}
% \email[]{pschmelc@physnet.uni-hamburg.de}
\affiliation{Zentrum f\"ur Optische Quantentechnologien, Universit\"{a}t Hamburg, Luruper Chaussee 149, 22761 Hamburg, Germany}
\affiliation{Hamburg Centre for Ultrafast Imaging, Universit\"{a}t Hamburg,
Luruper Chaussee 149, 22761 Hamburg, Germany}

\date{\today}

\begin{abstract}
The parity and Bloch theorems are generalized to the case of broken
global symmetry. Local inversion or translation symmetries are shown
to yield invariant currents that characterize wave propagation.
These currents map the wave function from an arbitrary spatial domain to
any symmetry-related domain. Our approach addresses any combination of
local symmetries, thus applying in particular to acoustic, optical
and matter waves. Nonvanishing values of the invariant currents provide a systematic
pathway to the breaking of discrete global symmetries.
\end{abstract}

\pacs{}

\maketitle

{\it Introduction.}---The use of symmetries in the description of nature constitutes the backbone of physics since its dawn, owing to their fundamental role in the theoretical treatment of any system.
From this point of view, important phenomenological properties of physical systems are often explained through a model with global symmetry. 
A prominent case are potentials with global inversion symmetry, leading to eigenstate wave functions with even or odd parity \cite{Zettili2009}, which we refer to as the `parity theorem', with far reaching consequences for the system's response.
Another particular case are spatially periodic potentials used in the description of atomic crystals.
The band structure of electronic states is here deduced from global translational symmetry on the basis of Bloch's theorem \cite{Bloch1929}, which relates their wave amplitudes under translation.
The global validity of a symmetry is, however, an idealized scenario which does not apply to many physical systems in reality.
Indeed, the multitude of structures appearing in nature is tightly related to physical mechanisms breaking the global or local presence of a symmetry, a process which usually leads to the emergence of new symmetries, possibly at different scales.
For artificially designed systems, broken global symmetry may even be dictated by technological constraints and functionality.
In practice one thus often deals with extended systems containing domains characterized locally by a certain symmetry, as is the case, e.g., for large molecules \cite{Pascal2001,Grzeskowiak1993}, quasi-crystals \cite{Lifshitz1996,Chen2012} self-organized, pattern-forming systems \cite{pattern} and even (partially) disordered matter \cite{Wochner2009}.
Despite this widespread situation of different local symmetries coexisting at different spatial scales, implied by the breaking of a global symmetry, a rigorous and functional theoretical treatment of such systems is missing.

A step in this direction was recently performed by introducing the concept of local parity \cite{Kalozoumis2013a} for the description of systems which can be decomposed into mirror-symmetric units. 
It was demonstrated how this local symmetry decomposition relates to spectral features such as perfect transmission resonances in aperiodic setups.
The origin of perfect transmission in such systems had yet been an unresolved issue, attributed to unknown `hidden'~\cite{Nava2009} or `internal'~\cite{Huang2001} symmetries.  
Within the local parity approach, an unambiguous symmetry-based classification of scattering states has been established \cite{Kalozoumis2013b}, thereby elucidating the link between perfect transmission and spatial symmetry. 
Developed within a generic framework of stationary wave mechanics, this treatment applies equally to any form of wave transmission as it occurs, e.g., in quantum devices \cite{Kalozoumis2013a}, classical optics \cite{Vardeny2013,Macia2012} or acoustic setups \cite{Hennion2013}.

The natural step beyond a classification of scattering states is to seek the exact way in which the remnants of a symmetry, which is broken globally, still determine the structure of a scattered wave.
Considering scattering in one dimension, there are two discrete symmetry transformations which can be rendered local, inversion through a given point and translation by a given length.
The question thus becomes:
What form, if any, will the associated parity and Bloch theorems acquire if global inversion or translation symmetry is broken?

\begin{figure}[t!]
    \begin{center}
      \includegraphics[width=.90\columnwidth]{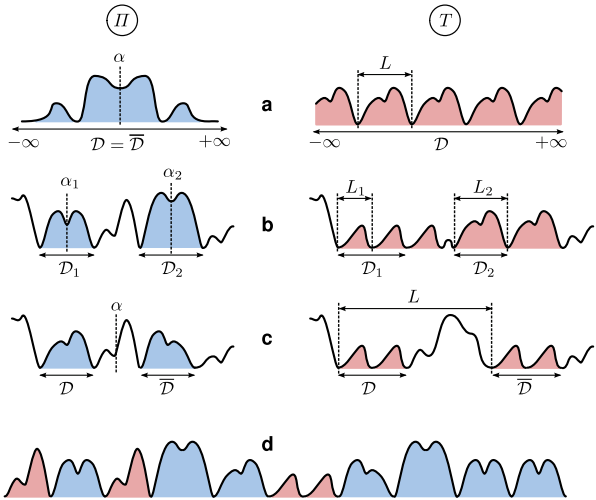}
    \end{center}
    \caption{
    (Color online) 
    Different types of symmetry under inversion ($\varPi$) through $\alpha$ or translation ($T$) by $L$, mapping a domain $\mathcal{D}$ to $\overline{\mathcal{D}}$: 
    (a) global symmetry, (b) non-gapped local symmetry, (c) gapped local symmetry, (d) complete local symmetry.
    }
  \label{fig}
\end{figure}

In the present Letter we answer this question by developing a theoretical framework for wave scattering in one-dimensional systems with inversion or translational symmetry within arbitrary (finite or infinite, connected or disconnected) spatial domains.
Employing a generic wave-mechanical setting, we first derive symmetry-induced, non-local currents which are spatially invariant within the domains where the corresponding symmetry is obeyed. 
We then demonstrate how these invariants can be used to explicitly map the solution of the associated wave equation across any local symmetry domain, in a unified way for inversion and translation.
This mapping is shown to generalize the parity and Bloch theorems, which are recovered in the limit of global symmetry.
The form and the consequences of this generalized mapping are finally demonstrated for different types of local symmetry (see Fig.~\ref{fig}).
A particular case studied is that of complete local symmetry, suggesting a class of structures composed exclusively of different connected segments, 
which are locally symmetric under inversion or translation.
 
{\it Invariant non-local currents.}---In a unified theoretical framework, one-dimensional stationary scattering of a complex wave field $\mathcal{A}(x)$ in absence of sinks and sources is described by the Helmholtz equation
%-----------------------------------------------------
\begin{equation}
\label{helmholtz} 
\mathcal{A}''(x)+U(x)\mathcal{A}(x)=0,
\end{equation}
%-----------------------------------------------------
where the prime denotes differentiation with respect to the spatial variable $x$.
Here $U(x)=\kappa^2(x)$ is a real function generated by an effective wave vector $\kappa(x)$, which describes the inhomogeneity of the medium where the wave propagates. 
For an electromagnetic wave of frequency $\omega$, $U(x)=\frac{\omega^2 n^2(x)}{c^2}$ is given by the refractive index $n(x)$ of the medium, and $\mathcal{A}(x)$ represents the complex amplitude of the electric field of the wave. 
For a matter wave, $U(x)=\frac{2 m}{\hbar^2}(E-V(x))$ is the scaled kinetic energy of a particle with mass $m$ and energy $E$ which moves in a potential $V(x)$ and is represented by the quantum wave function $\mathcal{A}(x)$.
Since we will focus in general on a scattering setting, we assume that the function $U(x)$, which we simply refer to as the `potential' in the following, is overall finite and asymptotically positive,  ${|U(x)|<\infty} ~~\forall ~x \in \mathbb{R}~; ~~ \lim_{x\to \pm \infty}U(x)>0.$
%-----------------------------------------------------

We now consider the spatial symmetry of $U(x)$ under the two elementary transformations in one dimension, inversion through a point $\alpha$ and translation by a length $L$.
To treat both cases within a common framework, let us define the linear transformation
%-----------------------------------------------------
\begin{equation}
\label{F}
F: ~ x \rightarrow \overline{x} = F(x) = \sigma x + \rho,
\end{equation}
%-----------------------------------------------------
whose type is determined by the parameters $\sigma$ and $\rho$ as follows:
%-----------------------------------------------------
\begin{eqnarray}
\sigma=-1,~ &\rho=2\alpha ~ \Rightarrow F = \varPi & ~: \mathrm{inversion~through}~\alpha \label{Pi}\\ 
\sigma=+1,~ &\rho=L ~ \Rightarrow  F = T & ~: \mathrm{translation ~by}~L
\label{T}
\end{eqnarray}
%-----------------------------------------------------
Assume now that the potential $U(x)$ obeys the symmetry 
%-----------------------------------------------------
\begin{equation}
\label{potsym}
U(x)=U(F(x))~~~~~\forall~~x \in \mathcal{D},
\end{equation}
%-----------------------------------------------------
for an arbitrary domain $\mathcal{D} \subseteq \mathbb{R}$.
If $\mathcal{D} = \mathbb{R}$, then the above symmetry is global, otherwise the symmetry is called local.
The latter can be of different types (see Fig.~\ref{fig}), to be discussed below.
The transformations $F$ generally map a domain $\mathcal{D}$ to a different one, $F(\mathcal{D})=\overline{\mathcal{D}} \neq \mathcal{D}$, the exception being inversion of a connected domain through its center $\alpha$ (as seen in Fig.~\ref{fig}(a) and (b) for $F=\varPi$).

The general local symmetry property of the potential can be exploited to construct locally spatial invariant quantities from the field $\mathcal{A}(x)$, as follows.
Evaluating Eq.~(\ref{helmholtz}) at a point $x$ and at its image $\overline{x} = F(x)$, we construct the difference $\mathcal{A}(\overline{x})\mathcal{A}''(x)-\mathcal{A}(x) \mathcal{A}''(\overline{x})$, which in general is non-zero and varies in $x$.
For a potential $U(x)$, however, which fulfills the considered symmetry property in Eq.~(\ref{potsym}), this expression vanishes for all $x \in \mathcal{D}$,
%-----------------------------------------------------
\begin{equation}
\label{conserv1}
\mathcal{A}(\overline{x})\mathcal{A}''(x)-\mathcal{A}(x) \mathcal{A}''(\overline{x})= 2iQ'(x)= 0,
\end{equation}
%-----------------------------------------------------
which implies that the complex quantity
%-----------------------------------------------------
\begin{equation}
\label{Q}
Q = \frac{1}{2i} \left[ \sigma \mathcal{A}(x)\mathcal{A}'(\overline{x}) - \mathcal{A}(\overline{x}) \mathcal{A}'(x) \right]
\end{equation}
%-----------------------------------------------------
is spatially invariant (constant in $x$) within the domain $\mathcal{D}$.
Here $\mathcal{A}'(\overline{x}) = \frac{d\mathcal{A}(x)}{dx}\rvert_{x=\overline{x}}$ denotes the derivative function evaluated at $\overline{x}=F(x)$, so that $\sigma$ distinguishes between the $Q$'s for inversion ($F=\varPi$) and translation ($F=T$).

The procedure can be repeated using the complex conjugate of Eq.~(\ref{helmholtz}) evaluated at $x$ to obtain another invariant in the same domain,
%-----------------------------------------------------
\begin{equation}
\label{Qtilde}
\widetilde{Q} = \frac{1}{2i} \left[ \sigma \mathcal{A}^*(x)\mathcal{A}'(\overline{x}) - \mathcal{A}(\overline{x}) \mathcal{A}'^*(x) \right].
\end{equation}
%-----------------------------------------------------
The invariants defined by Eqs.~(\ref{Q}) and (\ref{Qtilde}) have the form of a `non-local current' involving points connected by the corresponding symmetry transformation. 
Note that, because of their constancy in $\mathcal{D}$, $Q$ and $\widetilde{Q}$ can be evaluated from any chosen symmetry-related points $x$ and $\overline{x}$ in this domain.
The non-local currents within a symmetry domain $\mathcal{D}$ are related, for each of the two cases $F = \varPi,T$ ($\sigma = -1,+1$), by
%-----------------------------------------------------
\begin{equation}
\label{QJ}
|\widetilde{Q}|^2 -|Q|^2  = \sigma J^2,
\end{equation}
%-----------------------------------------------------
where $J$ is the usual 1D probability (or energy) current carried by the matter (or electromagnetic) wave, 
% -----------------------------------------------------
\begin{equation}
\label{J}
J = \frac{1}{2i}\left[\mathcal{A}^*(x)\mathcal{A}'(x) - \mathcal{A}(x)\mathcal{A}'^*(x)\right],
\end{equation}
% -----------------------------------------------------
which is locally defined and globally invariant in one dimension for the real potential $U(x)$.
The assumptions on the potential provided above guarantee that we have a propagating scattering state with $J \neq 0$.

{\it Generalized parity and Bloch theorems.}---We can now use the invariants $Q$, $\widetilde{Q}$ to  obtain a definite relation between the wave field $\mathcal{A}(x)$ and its image $\mathcal{A}(\overline{x})$ under a symmetry transformation.
As will be seen, this generalizes the usual parity and Bloch theorems to the case where global inversion and translation symmetry, respectively, is broken.
First, let us define an operator $\hat{O}_F$ which acts on $\mathcal{A}(x)$ by transforming its argument through $F=\varPi$ or $T$ as given in Eqs.~(\ref{F})-(\ref{T}), $\hat{O}_F \mathcal{A}(x) = \mathcal{A}(\overline{x}=F(x))$.
The image $\mathcal{A}(\overline{x})$ can then be solved for from the system of Eqs.~(\ref{Q}),~(\ref{Qtilde}), which finally yields
%-----------------------------------------------------
\begin{equation}
\label{genbpt}
\hat{O}_F \mathcal{A}(x) = \mathcal{A}(\overline{x})=\frac{1}{J} \left[ \widetilde{Q}\mathcal{A}(x) - Q \mathcal{A}^*(x) \right]
\end{equation}
%-----------------------------------------------------
for all $x \in \mathcal{D}$.
Equation (\ref{genbpt}), which remains valid in any type of domain $\mathcal{D}$ for states with $J \neq 0$, is a central result of the present work. 
It explicitly gives the image $\mathcal{A}(\overline{x})$ in the target domain $\overline{\mathcal{D}}$ as a linear combination of $\mathcal{A}(x)$ and its complex conjugate in $\mathcal{D}$, with the constant weights determined exclusively by $Q$ and $\widetilde{Q}$ (recall that $J$ is given from Eq.~(\ref{QJ})).
In other words, the invariant non-local currents $Q$ and $\widetilde{Q}$, induced by the generic symmetry of $U(x)$ in Eq.~(\ref{potsym}), provide the mapping between the field amplitudes at points related by this symmetry, regardless if the symmetry is global or not.
This generalized transformation of the field can therefore be identified as a remnant of symmetry in the case when it is globally broken.

{\it Globally symmetric potentials.}---To illustrate the strength of Eq.~(\ref{genbpt}) in a transparent way, consider the scenario where $\hat{O}_F$ commutes with the Helmholtz operator $\hat{\mathcal{H}}=\frac{d^2}{dx^2}+U(x)$, and thus $U(x)$ obeys global symmetry under $F$ with $\mathcal{D}=\mathbb{R}$ in Eq.~(\ref{potsym}), but the (asymptotic) boundary conditions on the field $\mathcal{A}(x)$ prevent it from being an eigenfunction of $\hat{O}_F$.
This is typically the case in a scattering situation, where incident waves are considered only on one side of the potential, thus breaking the $\varPi$-symmetry of the problem even if the potential is globally symmetric. 
Remarkably, within the present framework the benefit of an underlying symmetry is retained in the treatment of the problem: 
In spite of the asymptotic conditions breaking global symmetry, the quantities $Q$ and $\widetilde{Q}$ are constant in the entire space, relying on the global symmetry of $U(x)$, and Eq.~(\ref{genbpt}) still provides the mapping of the most general wave function under the action of $\hat{O}_F$. 

From Eq.~(\ref{genbpt}) it is now clearly seen that a nonvanishing $Q$ is a manifestation of broken global symmetry under the discrete transformation $F$.
This leads us to the recovery of the usual parity and Bloch theorems for globally $\varPi$- and $T$-symmetric systems by noting that, when $Q=0$, the field $\mathcal{A}(x)$ becomes an eigenfunction of $\hat{O}_{F=\varPi,T}$:
%-----------------------------------------------------
\begin{equation}
\label{eigenO}
\hat{O}_F \mathcal{A}(x) = \frac{\widetilde{Q}}{J}\mathcal{A}(x) \equiv \lambda_F \mathcal{A}(x).
\end{equation}
%-----------------------------------------------------
From Eq.~(\ref{QJ}) we also have that $|\frac{\widetilde{Q}}{J}| = 1$, so that any eigenvalue of $\hat{O}_F$ is restricted to the unit circle, $\lambda_F = e^{i\theta_F}$, as should be the case for the inversion ($\hat{O}_{\varPi}$) or translation ($\hat{O}_T$) operator.
Note here that $|\lambda_{\varPi}|$ cannot be determined from Eq.~(\ref{QJ}) with $Q=0$ and $\sigma=-1$, but shown below to be unity. 
Thus, with global $Q=0$, Eq.~(\ref{genbpt}) reduces to the form of the usual parity or Bloch theorem.
Indeed, let us distinguish here between the two cases $\varPi$ and $T$ and inspect them within the present framework of spatial invariants.

(i) For inversion $\varPi$, we immediately get $\lambda_{\varPi} = \pm 1$ (or $\theta_{\varPi}=0,\pi$) from Eq.~(\ref{eigenO}), since necessarily $\hat{O}_{\varPi}^2$ is the identity operator.
Alternatively, we can evaluate Eq.~(\ref{eigenO}) and its first derivative at the center of inversion $x=\alpha$, which gives $\mathcal{A}(\alpha)=\lambda_{\varPi}\mathcal{A}(\alpha)$ and $\mathcal{A}'(\alpha)=-\lambda_{\varPi}\mathcal{A}'(\alpha)$.
Then $\lambda_{\varPi} = +1$ (with $\mathcal{A}'(\alpha)=0$) or $-1$ (with $\mathcal{A}(\alpha)=0$), corresponding to even and odd eigenfunctions, respectively, which constitutes the parity theorem.
Setting $Q=0$ and $\sigma = -1$ in Eq.~(\ref{QJ}) implies that $\widetilde{Q}=J=0$, which represents a special case in which the generalized symmetry-induced mapping, Eq.~(\ref{genbpt}), obviously collapses.
However, the mapping is still trivially provided by the global parity of the field $\mathcal{A}(x)$.  
Equation~(\ref{eigenO}) is here realized either for bound states (which are, however, not supported by the considered potential) with symmetric boundary conditions, or for scattering states with symmetrically incoming waves from both sides of the potential $U(x)$ such that $J=0$. 
The latter are the zero-current states discussed in Ref.~\cite{Kalozoumis2013a}.

(ii) For translation $T$, Eq.~(\ref{potsym}) implies global periodicity of the potential, so that $U(x) =U(x+nL)$ for any $n$ $\in$ $\mathbb{Z}$. 
Defining a different $\widetilde{Q}_n$ for each $n$, the corresponding translation through $\hat{O}_T$ can be performed equivalently either in a single translation by $nL$ or in $n$ successive translations by $L$:
$\mathcal{A}(x+nL) = (\widetilde{Q}_n/J)\mathcal{A}(x) = (\widetilde{Q}_1/J)^n\mathcal{A}(x)$.
This yields that the phase of $\lambda_T$ in Eq.~(\ref{eigenO}) increases linearly in $L$, $\arg{(\widetilde{Q}_n/J)} = n \arg{(\widetilde{Q}_1/J)} = kL$, where $k$ is a constant of inverse length dimension.
Bloch's theorem is thus recovered by identifying $k$ as the crystal momentum in an infinitely periodic system.
In this limit, the phase $\theta_{\widetilde{Q}}$ of $\widetilde{Q}$ becomes the Bloch phase up to multiples of $\pi$, $\widetilde{Q} = \pm|J| e^{i \theta_{\widetilde{Q}}} = \pm|J| e^{ikL}$, according to the direction of the current $J$.

{\it Locally symmetric potentials.}---
Having treated the case where global symmetry is broken through asymptotic conditions, we now turn to the intriguing case where the potential $U(x)$ itself is globally non-symmetric under the transformations $F = \varPi, T$.
Since $\hat{O}_{F}$ does not commute with $\hat{H}$ in this case, one cannot seek stationary eigenstates in analogy to Eq.~(\ref{eigenO}), unless the action of $\hat{O}_{F}$ is spatially partitioned, as was done for $\varPi$-symmetry in Ref.~\cite{Kalozoumis2013a}.
We here choose to analyze the local symmetry structure of the system in terms of the invariants $Q$ and $\widetilde{Q}$, which explicitly relate the symmetry information to the scattering wave amplitudes via Eq.~(\ref{genbpt}).

In the extreme case of total $F$-symmetry breaking, there is no remnant of symmetry present in $U(x)$, and therefore also no domain $\mathcal{D}$ with constant $Q$.
Although one can still define spatial functions $Q(x)$,~$\widetilde{Q}(x)$ as in Eqs.~(\ref{Q}),~(\ref{Qtilde}), leading to Eq.~(\ref{genbpt}), their non-constancy brings no advantage to the representation of the scattering problem.
If local symmetry is present in $U(x)$, we can distinguish the following characteristic cases, which are sketched in Fig.~\ref{fig}.

(i) Non-gapped local symmetries:
In this case, a symmetry domain $\mathcal{D}$ either coincides, overlaps, or connects with its image $\overline{\mathcal{D}}=F(\mathcal{D})$ (that is, $\overline{\mathcal{D}} \cup \mathcal{D}$ is connected).
There can exist one or many such domains of symmetry along the total potential landscape $U(x)$, in general with  non-symmetric parts in between; see Fig.~\ref{fig}(b).
To each local symmetry domain $\mathcal{D}_i$ ($i=1,2,..,N$) a pair of spatially invariant (within $\mathcal{D}_i$) non-local currents $Q_i$ and $\widetilde{Q}_i$ is associated, whose different values change with the energy $E$ (or frequency $\omega$) of the field.
Within each domain, Eq.~(\ref{genbpt}) spatially maps $\mathcal{A}(x)$ to its image under the corresponding symmetry transformation:
For local $\varPi$-symmetry, the field in one half of ${\mathcal{D}}_i$ is determined from the field in the other half through the $Q_i$ and $\widetilde{Q}_i$, which can be evaluated at the center of inversion $\alpha_i$.
For local $T$-symmetry, the field within the first interval of length $L$ in ${\mathcal{D}}_i$ successively determines its images along a locally periodic part of $U(x)$, 
and the pair of invariants can be evaluated at the corresponding boundaries.

(ii) Gapped local symmetries:
In this case, a domain $\mathcal{D}$ has no overlap with its symmetry-related image ($\overline{\mathcal{D}} \cap \mathcal{D} = \varnothing$).
In other words, there is a gap between them in configuration space, within which $U(x)$ can obey another or no symmetry; see Fig.~\ref{fig}(c).
This can of course occur if the source domain $\mathcal{D}$ is not connected, that is, already has gaps in it.
For connected $\mathcal{D}$, gapped $\varPi$-symmetry occurs if the inversion point lies outside the associated domain ($\alpha \notin \mathcal{D}$), and gapped $T$-symmetry if the translation length $L$ exceeds the size of $\mathcal{D}$. 
Interestingly, once $Q$ and $\widetilde{Q}$ have been evaluated from a pair of symmetry-connected points, Eq.~(\ref{genbpt}) maps the wave function from one part of the potential to a remote part, although there is an arbitrary field variation in the intervening gap. 

(iii) Complete local symmetry (CLS): 
An appealing situation occurs when the potential $U(x)$ can be completely decomposed into locally symmetric domains, each one characterized by a remnant of the broken symmetry; see Fig.~\ref{fig}(d). 
This can be realized by attached domains of non-gapped local $\varPi$- or $T$-symmetry, of a single kind or mixed.
Also gapped local symmetries can be part of a such a structure, with the gaps between their source ($\mathcal{D}_i$) and image ($\overline{\mathcal{D}}_i$) domains all filled in either by non-gapped local symmetry domains or by the source or image domain of other gapped symmetries.
The latter case can then result in multiply intertwined symmetry domains, in a way that makes even the presence of a local symmetry structure far from evident.
The non-local currents $Q$,~$\widetilde{Q}$ can then be utilized as a detection tool for local symmetry:
Calculating them for every pair $(x,\bar{x})$, using different $\alpha$ or $L$, their constancy would reveal underlying symmetry domains, if present.

Consider, as an example, a locally $\varPi$-symmetric potential $U(x)=\sum_{i=1}^{N} U_i(x)$ defined on successive non-overlapping domains $\mathcal{D}_i$ with centers $\alpha_i$, such that $U_i(2 \alpha_i -x) = U_i (x)$ for $x \in \mathcal{D}_i$ and $U_i(x) = 0$ for $x \notin \mathcal{D}_i$ \cite{Kalozoumis2013a,Kalozoumis2013b}.
As before, Eq.~(\ref{genbpt}) gives the wave function in one half of each domain from that in the other half through the local invariants $Q_i$ and $\widetilde{Q}_i$.
Our approach here reveals an interesting relation between global and complete local $\varPi$-symmetry of $U(x)$.
In total, both cases map the field in one half of the entire configuration space to the other half, though the domains of the source $\mathcal{A}(x)$ are topologically different:
a connected domain for global symmetry, and a disconnected domain for CLS.
In addition, the relation of the $Q_i$,~$\widetilde{Q}_i$ to the globally invariant current $J$ provides a constraint between their magnitudes in all different domains,
% -----------------------------------------------------
\begin{equation}
\label{conQ}
\frac{|Q_{i+1}|^2 - |\widetilde{Q}_{i+1}|^2}{|Q_i|^2 - |\widetilde{Q}_i|^2} = 1, ~~i=1,2,...,N-1.
\end{equation}
% -----------------------------------------------------
The local invariants of all domains can then be combined to form two overall, piecewise constant functions $Q_c(x)$ and $\widetilde{Q}_c(x)$, which characterize the CLS of the structure at a given energy (or frequency) of the field.
This suggests a characteristic class of CLS material structures, which generalize the notion of periodic or aperiodic crystals.

{\it Conclusions.}---Considering structures which are locally symmetric with respect to inversion or translation, we have shown how a pair of non-local currents, $Q$ and $\widetilde{Q}$, characterize generic wave propagation within arbitrary symmetry domains. 
In particular, these invariant currents comprise the information necessary to map the wave function from a spatial subdomain to any symmetry-related subdomain.
In this sense our theoretical framework generalizes the parity and Bloch theorems from global to local inversion and translation symmetries.
Both invariant currents represent a (local) remnant of the corresponding global symmetry, and nonvanishing $Q$ is identified as the key to the breaking of global symmetry.
Although we focus here on one spatial dimension, the appealing compact form of our results suggest that the formalism can be extended to two or three dimensions.
Further, due to the general wave mechanical framework our approach applies equally to quantum or classical wave scattering and should therefore have a diversity of potential applications in, e.g., nanoelectronic devices,
photonic crystals or acoustic setups.
Structures consisting exclusively of locally symmetric building blocks should here be of particular interest.
Due to the connection of the introduced invariant currents to global symmetry breaking and to scattering in locally symmetric systems, they are expected to trigger a deeper understanding--and the finding of novel structures--of wave propagation phenomena in complex media. 
  
{\it Acknowledgments.}---We are grateful to V. Zampetakis and M. Diakonou for helpful discussions. 
P. A. K. acknowledges a scholarship by the  European Union (European Social Fund - ESF) and Greek national funds through the Operational Program ``Education and Lifelong Learning" of the National Strategic Reference Framework (NSRF) - Research Funding Program: Heracleitus II. Investing in knowledge society through the European Social Fund.
Financial support by the Deutsche Forschungsgemeinschaft (DFG) through the excellence cluster 'The Hamburg Centre for Ultrafast Imaging - Structure, Dynamics and Control
of Matter on the Atomic Scale' in the framework of a collaborative visit is gratefully acknowledged.

\end{document}